\documentclass{ws-ijmpcs}

\usepackage{amsmath,amssymb,epsfig,fleqn,url,psboxit,color,here,graphics,graphicx,rotating,multirow,wrapfig}

\usepackage{hyperref}
\hypersetup{
    colorlinks,
    citecolor=blue,
    filecolor=blue,
    linkcolor=blue,
    urlcolor=blue
}

\newcommand{\slim}{\mskip 1.5mu}              % small space in math

\newcommand{\phiH}{\phi _h}
\newcommand{\phiS}{\phi _S}

\begin{document}

\markboth{Bakur Parsamyan}{Multidimensional TSAs at COMPASS}

\catchline{}{}{}{}{}
%
%%%%%%%%%%%%%%%%%%%%%%%%%%%%%%%%%%%%%%%%%%%%%%%%%%%%%%%%%%%%%%%%%%%%

\title{Transverse spin azimuthal asymmetries in SIDIS at COMPASS: Multidimensional analysis}

\author{Bakur Parsamyan}
\address{
ICTP - Strada Costiera 11, 34151 Trieste, Italy
\\
INFN sezione di Trieste - Via Valerio 2, 34127 Trieste, Italy
\\
Universit\`a di Torino \and INFN sezione di Torino -
Via P. Giuria 1, 10125 Torino, Italy
\\
bakur.parsamyan@cern.ch}

\maketitle

%\begin{history}
%\received{Day Month Year}
%\revised{Day Month Year}
%\published{Day Month Year}
%\end{history}

\begin{abstract}
COMPASS is a high-energy physics experiment operating at the SPS at CERN. Wide physics program of the experiment comprises study of hadron structure and spectroscopy with high energy muon and hadrons beams. As for the muon-program, one of the important objectives of the COMPASS experiment is the exploration of the transverse spin structure of the nucleon via spin (in)dependent azimuthal asymmetries in single-hadron production in deep inelastic scattering of polarized leptons off transversely polarized target. For this purpose a series of measurements were made in COMPASS, using 160 GeV/c longitudinally polarized muon beam and transversely polarized $^{6}LiD$ (in 2002, 2003 and 2004) and $NH_{3}$ (in 2007 and 2010) targets.	
	The experimental results obtained by COMPASS for unpolarized target azimuthal asymmetries, Sivers and Collins effects and other azimuthal observables play an important role in the general understanding of the three-dimensional nature of the nucleon. Giving access to the entire “twsit-2” set of transverse momentum dependent parton distribution functions and fragmentation functions COMPASS data triggers constant theoretical interest and is being widely used in phenomenological analyses and global data fits.	
	In this review main focus is given to the very recent results obtained by the COMPASS collaboration from first ever multi-dimensional extraction of transverse spin asymmetries.
%\footnote{At the SPIN-2014 conference these results have been presented for the first time}.

\end{abstract}

\ccode{PACS numbers: 13.60.-r; 13.60.Hb; 13.88.+e; 14.20.Dh; 14.65.-q.}
\section{Introduction}	
Using standard notations the cross-section for the lepton off transversely polarized nucleon SIDIS processes (in a single photon exchange approximation) can be written in a following model-independent way\cite{Kotzinian:1994dv}\cdash\cite{Diehl:2005pc}:
{\footnotesize
\begin{eqnarray}\nonumber
&& \hspace*{-1.4cm}\frac{{d\sigma }}{{dxdydzp_{T}^{h}dp_{T}^{h}d{\phiH}d\phiS }} = 2\left[ {\frac{\alpha }{{xy{Q^2}}}\frac{{{y^2}}}{{2\left( {1 - \varepsilon } \right)}}\left( {1 + \frac{{{\gamma ^2}}}{{2x}}} \right)} \right]\left( {{F_{UU,T}} + \varepsilon {F_{UU,L}}} \right) \\ \nonumber
%
%&&\hspace*{-1.4cm} \times\Bigg\{ 1 + unpolarized\,asymmetries +\,{S_{L}}\Big[ longitudinal\,asymmetries \Big]\\ \nonumber
%
&&\hspace*{-1.4cm} \times\Bigg\{ 1 + \sqrt {2\varepsilon \left( {1 + \varepsilon } \right)} \textcolor[rgb]{0.00,0.07,1.00}{A_{UU}^{\cos {\phi _h}}}\cos {\phiH} + \varepsilon \textcolor[rgb]{1.00,0.00,0.00}{A_{UU}^{\cos 2{\phi _h}}}\cos \left( {2{\phiH}} \right) + \lambda \sqrt {2\varepsilon \left( {1 - \varepsilon } \right)} \textcolor[rgb]{0.00,0.07,1.00}{A_{LU}^{\sin {\phi _h}}}\sin {\phiH}\\ \nonumber
%&&\hspace*{-1.1cm}+\,{S_{L}}\Big[ {\sqrt {2\varepsilon \left( {1 + \varepsilon } \right)} \sin {\phiH}A_{UL}^{\sin {\phiH}} + \varepsilon \sin \left( {2{\phiH}} \right)A_{UL}^{\sin \left( {2{\phiH}} \right)}} \Big]\\ \nonumber
%&&\hspace*{-1.1cm}+\,{S_{L}}\lambda \Big[ {\sqrt {1 - {\varepsilon ^2}} {A_{LL}} + \sqrt {2\varepsilon \left( {1 - \varepsilon } \right)} \cos {\phiH}A_{LL}^{\cos {\phiH}}} \Big]\\ \nonumber
&&\hspace*{-1.1cm}+\,{{S}_{T}}\Big[\textcolor[rgb]{1.00,0.00,0.00}{A_{UT}^{\sin \left( {{\phiH} - {\phiS}} \right)}}\sin \left( {{\phiH} - {\phiS}} \right) + \varepsilon \textcolor[rgb]{1.00,0.00,0.00}{A_{UT}^{\sin \left( {{\phiH} + {\phiS}} \right)}}\sin \left( {{\phiH} + {\phiS}} \right) + \varepsilon \textcolor[rgb]{1.00,0.00,0.00}{A_{UT}^{\sin \left( {3{\phiH} - {\phiS}} \right)}}\sin \left( {3{\phiH} - {\phiS}} \right)\\ \nonumber
&&\hspace*{+0.1cm}+\,\sqrt {2\varepsilon \left( {1 + \varepsilon } \right)} \textcolor[rgb]{0.00,0.07,1.00}{A_{UT}^{\sin {\phiS}}}\sin {\phiS} + \sqrt {2\varepsilon \left( {1 + \varepsilon } \right)} \textcolor[rgb]{0.00,0.07,1.00}{A_{UT}^{\sin \left( {2{\phiH} - {\phiS}} \right)}}\sin \left( {2{\phiH} - {\phiS}} \right)\Big]\\ \nonumber
&&\hspace*{-1.1cm}+\,{{S}_{T}}\lambda \Big[\sqrt {\left( {1 - {\varepsilon ^2}} \right)} \textcolor[rgb]{1.00,0.00,0.00}{A_{LT}^{\cos \left( {{\phiH} - {\phiS}} \right)}}\cos \left( {{\phiH} - {\phiS}} \right)\\
&&\hspace*{+0.1cm}+\,\sqrt {2\varepsilon \left( {1 - \varepsilon } \right)} \textcolor[rgb]{0.00,0.07,1.00}{A_{LT}^{\cos {\phiS}}}\cos {\phiS} + \sqrt {2\varepsilon \left( {1 - \varepsilon } \right)} \textcolor[rgb]{0.00,0.07,1.00}{A_{LT}^{\cos \left( {2{\phiH} - {\phiS}} \right)}}\cos \left( {2{\phiH} - {\phiS}} \right)\Big]\Bigg\}
\label{eq:SIDIS}
\end{eqnarray}
\\
}
where ratio of longitudinal and transverse photon fluxes is given as $\varepsilon = (1-y -\frac{1}{4}\slim \gamma^2 y^2)/(1-y +\frac{1}{2}\slim y^2 +\frac{1}{4}\slim \gamma^2 y^2)$; $\gamma = 2 M x/Q$.
 Target transverse polarization (${S}_{T}$) dependent part of this general expression contains eight azimuthal modulations in the $\phi_h$ and $\phi_S$ azimuthal angles of the produced hadron and of the nucleon spin, correspondingly. Each modulation leads to a $A_{BT}^{w_i(\phiH, \phiS)}$ Transverse-Spin-dependent Asymmetry (TSA) defined as a ratio of the associated structure function $F_{BT}^{w_i(\phiH,\phiS)}$ to the azimuth-independent one ${F_{UU}}={{F_{UU,T}} + \varepsilon {F_{UU,L}}}$. Here the superscript of the asymmetry
indicates corresponding modulation, the first and the second subscripts - respective ("U"-unpolarized,"L"-longitudinal and
"T"-transverse) polarization of beam and target. Five amplitudes which depend only on ${S}_{T}$ are the target Single-Spin Asymmetries (SSA), the other three
which depend both on ${S}_{T}$ and $\lambda$ (beam longitudinal polarization) are known as Double-Spin Asymmetries (DSA).
%Amplitude of each modulation is scaled by a $\varepsilon$-dependent so-called depolarization factor where:
%

In the QCD parton model approach four (marked in red) out of eight transverse spin asymmetries have Leading Order (LO) or leading "twist" interpretation and are described by the convolutions of twist-two Transverse-Momentum-Dependent (TMD) Parton Distribution Functions (PDFs) and Fragmentation Functions (FFs)\cite{Kotzinian:1994dv}\cdash\cite{Mulders:1995dh}.
These are the famous $A_{UT}^{sin(\phi_h+\phi_S)}$ "Sivers" and $A_{UT}^{sin(\phi_h+\phi_S)}$ "Collins" effects \cite{Adolph:2012sn,Adolph:2012sp} and $A_{UT}^{\sin(3\phiH -\phiS )}$ single-spin asymmetry (related to $h_{1T}^{\perp\,q}$ ("pretzelosity") PDF\cite{Parsamyan:2014uda}\cdash\cite{Parsamyan:2007ju}) and $A_{LT}^{\cos (\phiH -\phiS )}$ DSA (related to $g_{1T}^q$ ("worm-gear") distribution function\cite{Parsamyan:2014uda}\cdash\cite{Parsamyan:2007ju,Kotzinian:2006dw,Anselmino:2006yc}).

Remaining four (marked in blue) asymmetries ($A_{UT}^{\sin (\phi _s )}$ and $A_{UT}^{\sin (2\phi _h -\phi _s )}$ SSAs and $A_{LT}^{\cos (\phi _s)}$ and $A_{LT}^{\cos (2\phi _h -\phi _s )}$ DSAs) are so-called "higher-twist" effects. Corresponding structure functions contain terms at sub-leading order in $Q^{-1}$ which involve a mixture of twist-two and induced by quark-gluon correlations twist-three parton distribution and fragmentation functions\cite{Bacchetta:2006tn,Mao:2014aoa,Mao:2014fma}.
However, applying wildly adopted so-called "Wandzura-Wilczek approximation" this higher twist objects can be simplified to twist-two level (see Refs.~\refcite{Bacchetta:2006tn,Mulders:1995dh} for more details).

%All listed TSAs were extracted from COMPASS experimental data collected on transversely polarized
%deuteron and proton targets (See Refs.~\refcite{Adolph:2012sn}--\refcite{Parsamyan:2007ju} and references therein).

In general, TSAs being convolutions of different TMD functions are known to be complex objects \textit{a priori} dependent on the choice of kinematical ranges and multidimensional kinematical phase-space.
Thus, ideally, asymmetries have to be extracted as multi-differential functions of kinematical variables in order to reveal the most complete multivariate dependence. In practice, available statistics often is too limited for such an ambitious approach and investigating dependence of the asymmetries on some specific kinematic variable one is forced to integrate over all the others.
Presently, one of the hottest topics in the field of spin-physics is the study of TMD evolution of various PDFs and FFs and related asymmetries. Different models predict from small up to quite large $\sim1/Q^2$ suppression of the QCD-evolution effects attempting to describe available experimental observations and make predictions for the future ones \cite{Aybat:2011ta,Echevarria:2014xaa,Sun:2013hua}. Additional experimental measurements exploring different $Q^2$ domains for fixed $x$-range are necessary to further constrain the theoretical models.
The work described in this review is a unique and first ever attempt to explore behaviour of TSAs in the multivariate kinematical environment.
For this purpose COMPASS experimental data was split into five different $Q^2$ ranges giving an opportunity to study asymmetries as a function of $Q^2$ at fixed bins of $x$. Additional variation of $z$ and $p_T$ cuts allows to deeper explore multi-dimensional behaviour of the TSAs and their TMD constituents.
\section{Analysis and COMPASS multi-dimensional concept}	
The analysis was carried out on COMPASS data collected in 2010 with transversely polarized proton data.
General, event selection procedure as well as asymmetry extraction and systematic uncertainty definition techniques applied for this analysis
are identical to those used for recent COMPASS results on Collins, Sivers and other TSAs \cite{Adolph:2012sn}\cdash\cite{Parsamyan:2007ju}.
%The eight target transverse spin dependent "raw" asymmetries are extracted simultaneously from the fit using extended unbinned maximum likelihood method and then are corrected for average depolarization factors ($\varepsilon$-depending factors in equation (\ref{eq:x_sec}) standing in front of the amplitudes), dilution factor and target and beam (only DSAs) polarizations evaluated in the given kinematical bin \cite{Adolph:2012sn}\cdash\cite{Parsamyan:2007ju}.
%
%
General sample is defined by the following standard DIS cuts: $Q^2>1$ $(GeV/c)^2$, $0.003<x<0.7$ and $0.1 <y < 0.9$ and two more \textit{hadronic} selections: $p_T>0.1$ GeV/c and $z>0.1$.

\begin{figure}[h]
\includegraphics[width=6.3cm]{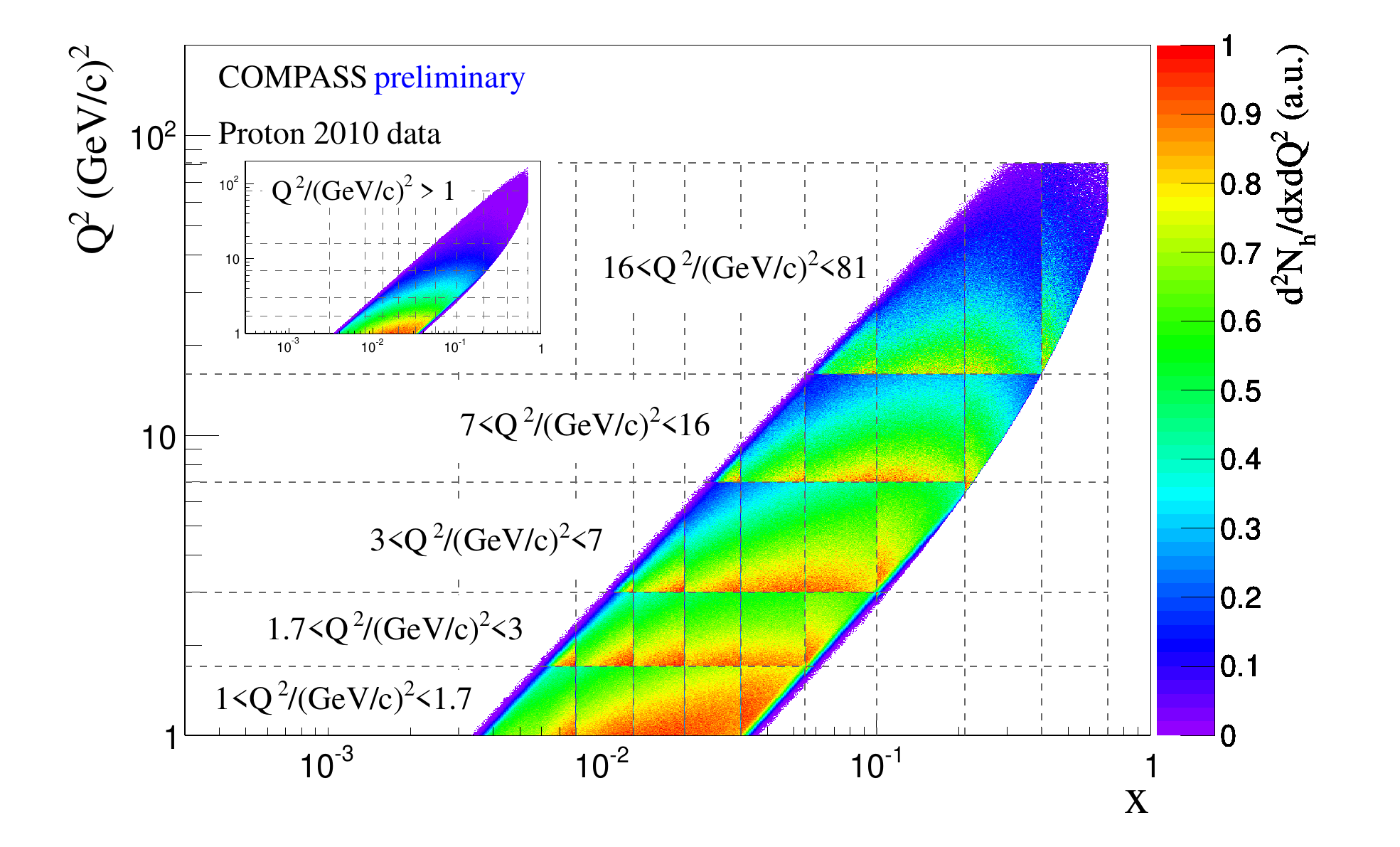}
\includegraphics[width=6.3cm]{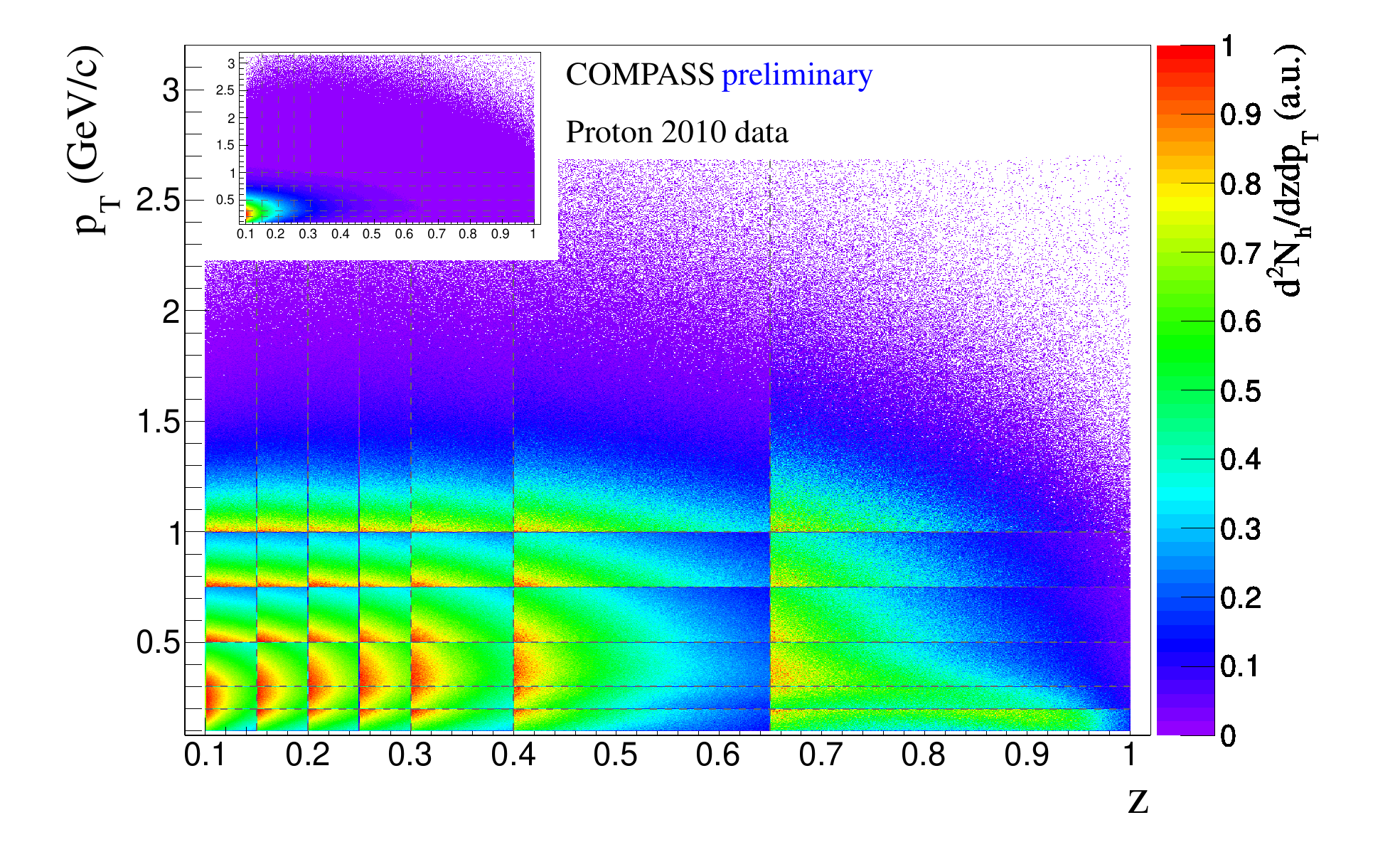}
%\vspace*{3pt}
\caption{COMPASS $x:Q^2$ (right) and $z$:$p_T$ (left) phase space coverage. \label{f1}}
\end{figure}

In order to study possible $Q^2$-dependence the $x$:$Q^2$ phase-space covered by COMPASS experimental data has been divided into $5\times9$ two-dimensional grid (see right plot in Fig.~\ref{f1}). Selected five $Q^2$-ranges are the following ones: $Q^{2}/(GeV/c)^2$ $\in$ $[1;1.7],[1.7;3],[3;7],[7;16],[16;81]$
\footnote{$16<Q^{2}/(GeV/c)^2<81$ selection repeats the definition of the so-called "high-mass" range: most promising domain for future COMPASS--Drell-Yan TSA-analyses \cite{Parsamyan:2014uda,Parsamyan:2015cfa}.}. In addition, each of this samples has been divided into five $z$ and five $p_T$ (GeV/c) sub-ranges defined as follows:\\
$z>0.1$, $z>0.2$, $0.1<z<0.2$, $0.2<z<0.4$ and $0.4<z<1.0$\\
$p_T>0.1$, $0.1<p_T<0.75$, $0.1<p_T<0.3$, $0.3<p_T<0.75$ and $p_T>0.75$.
Using various combinations of aforementioned cuts and ranges, asymmetries have been extracted for following "3D" and "4D" configurations: 1) $x$-dependence in $Q^2$-$z$ and $Q^2$-$p_T$ grids. 2) $Q^2$-dependence in $x$-$z$ and $x$-$p_T$ grids. 3) $Q^2$- (or $x$-) dependence in $x$-$p_T$ (or $Q^2$-$p_T$) grids for different choices of $z$-cuts.

Another approach was used to focus on $z$- and $p_T$-dependences in different $x$-ranges. For this study the two-dimensional $z$:$p_T$ phase-space has been divided into $7\times6$ grid as it is demonstrated in left plot in Fig.~\ref{f1}. Selecting in addition three $x$-bins: $0.003<x<0.7$, $0.003<x<0.032$, $0.032<x<0.7$ asymmetries have been extracted in "3D: $x$-$z$-$p_T$" grid.
In the next section several examples of COMPASS preliminary results obtained for multi-dimensional target transverse spin dependent azimuthal asymmetries are presented.
\section{Results}
\begin{figure}[H]
\includegraphics[width=12.8cm]{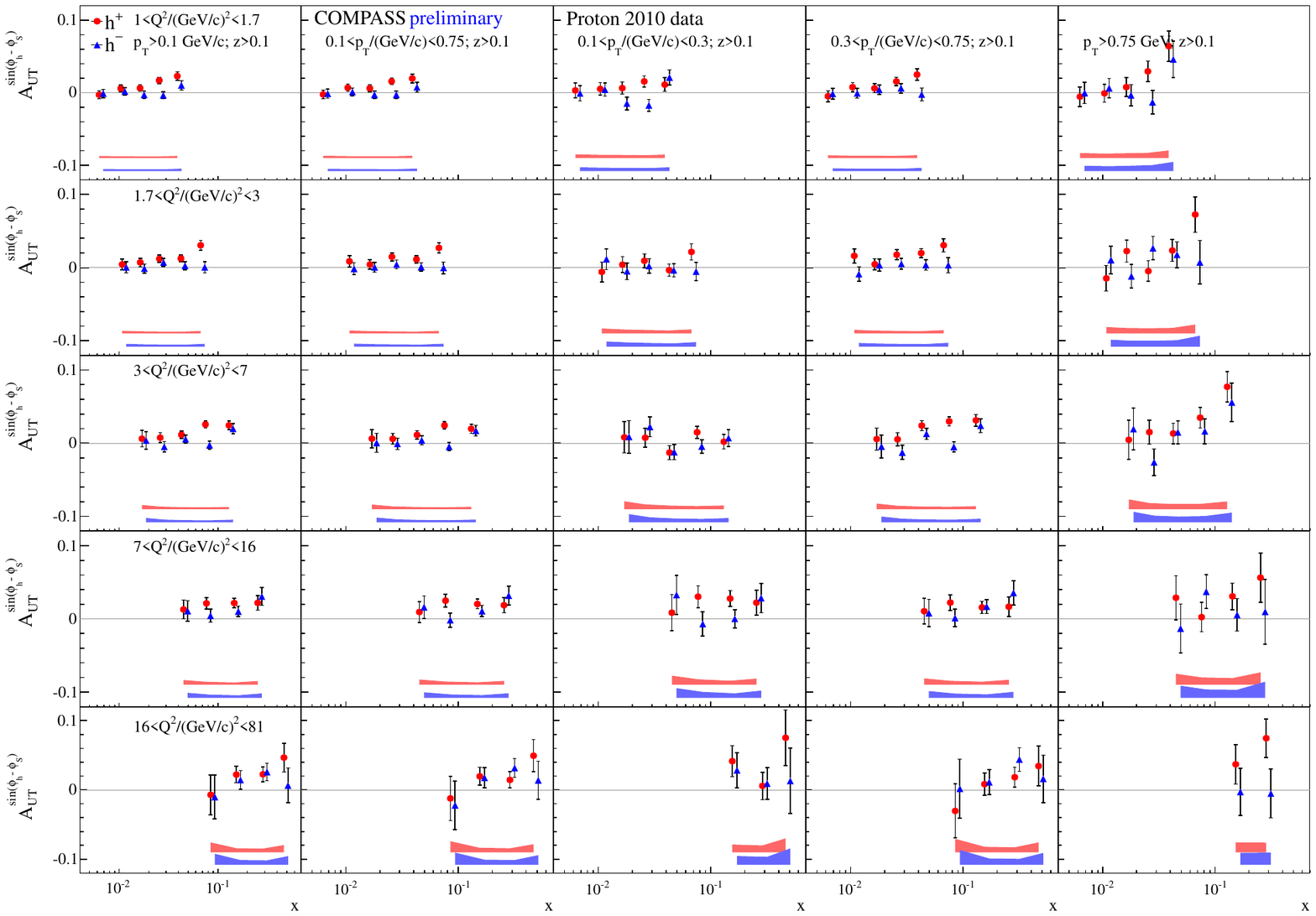}
\includegraphics[width=12.8cm]{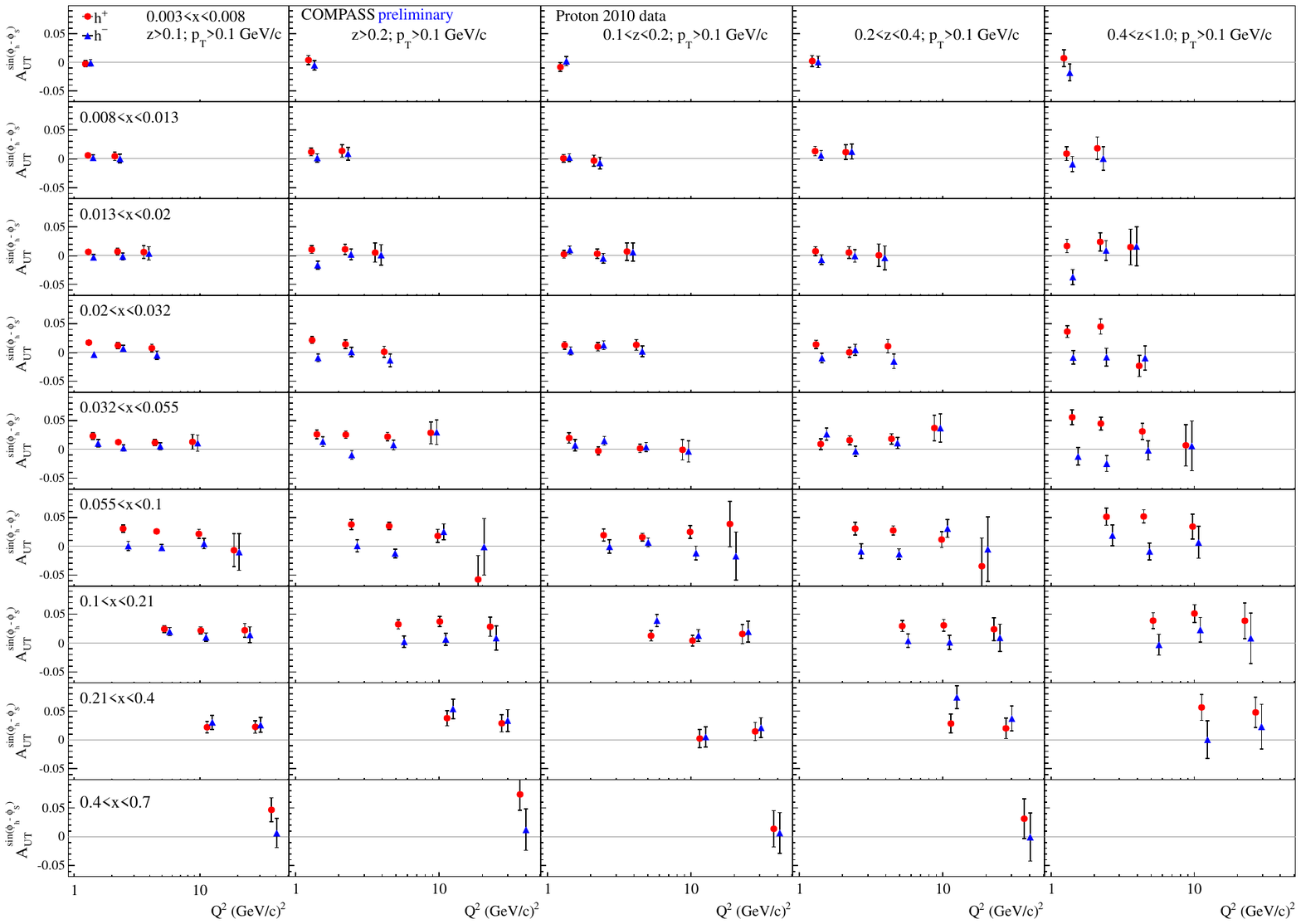}
\vspace*{8pt}
\caption{Sivers asymmetry in "3D": $Q^2$-$p_T$-$x$ (top) and $x$-$z$-$Q^2$ (bottom). \label{f2}}
\end{figure}

As an example of "3D" Sivers effect, two extracted configurations are quoted in the Fig.~\ref{f2}. In the top plot $x$-dependence of the asymmetry for positive and negative hadrons is shown in cells of two-dimensional grid of $Q^2$ and $p_T$ exploring the "3D: $Q^2$-$p_T$-$x$" behaviour. Sizable Sivers asymmetry is observed for positive hadrons tending to increase with $p_T$, while for negative hadrons there are some indications for a positive signal at relatively large $x$ and $Q^2$.
The bottom plot draws the $Q^2$ dependence in a "3D: $x$-$z$-$Q^2$" configuration and serves as a direct input for TMD-evolution related studies. In fact, in several x-bins there are some hints for possible $Q^2$-dependence for positive hadrons (decrease) more evident at large $z$ bins.

\begin{figure}[H]
\includegraphics[width=12.8cm]{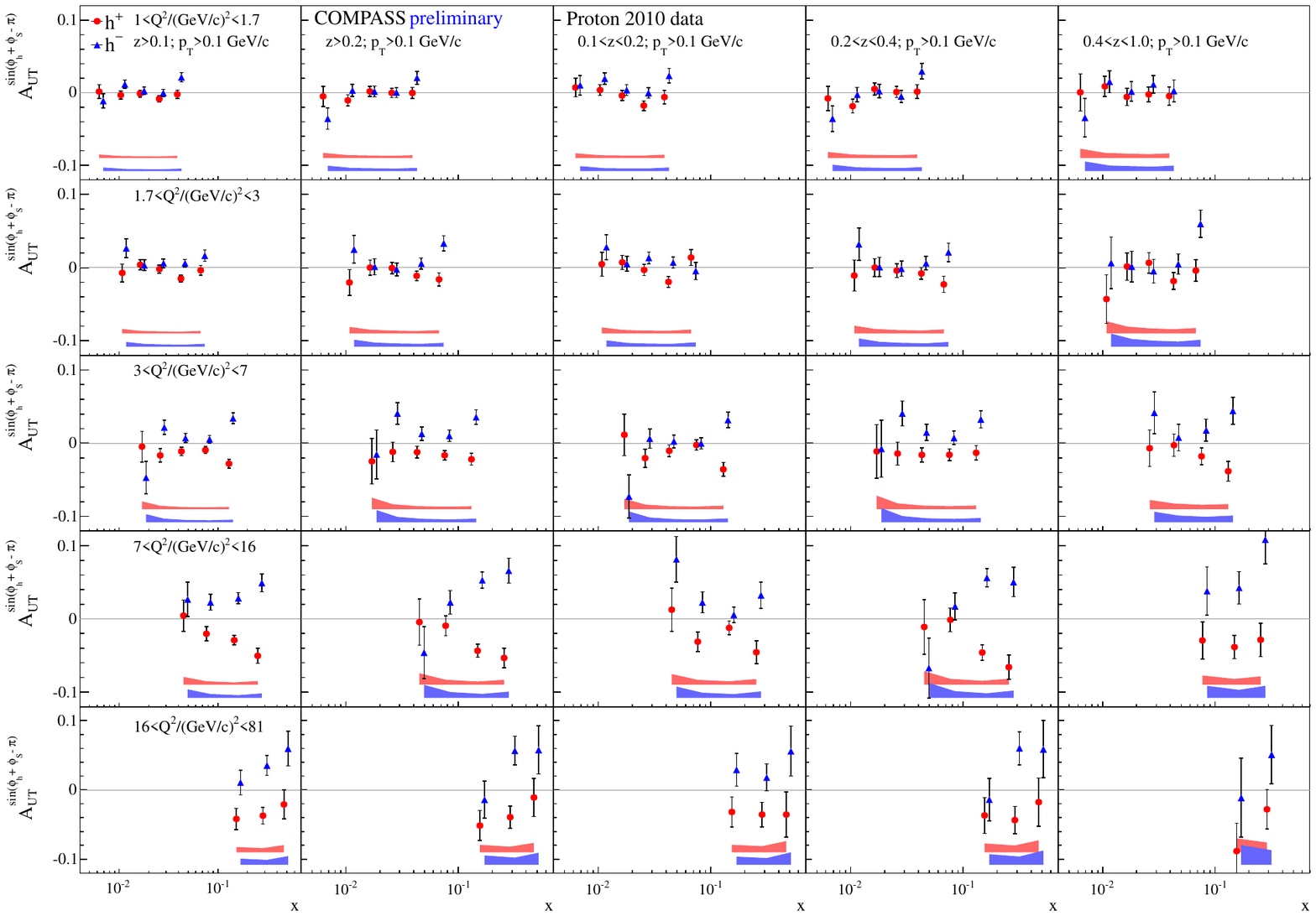}
\includegraphics[width=12.8cm]{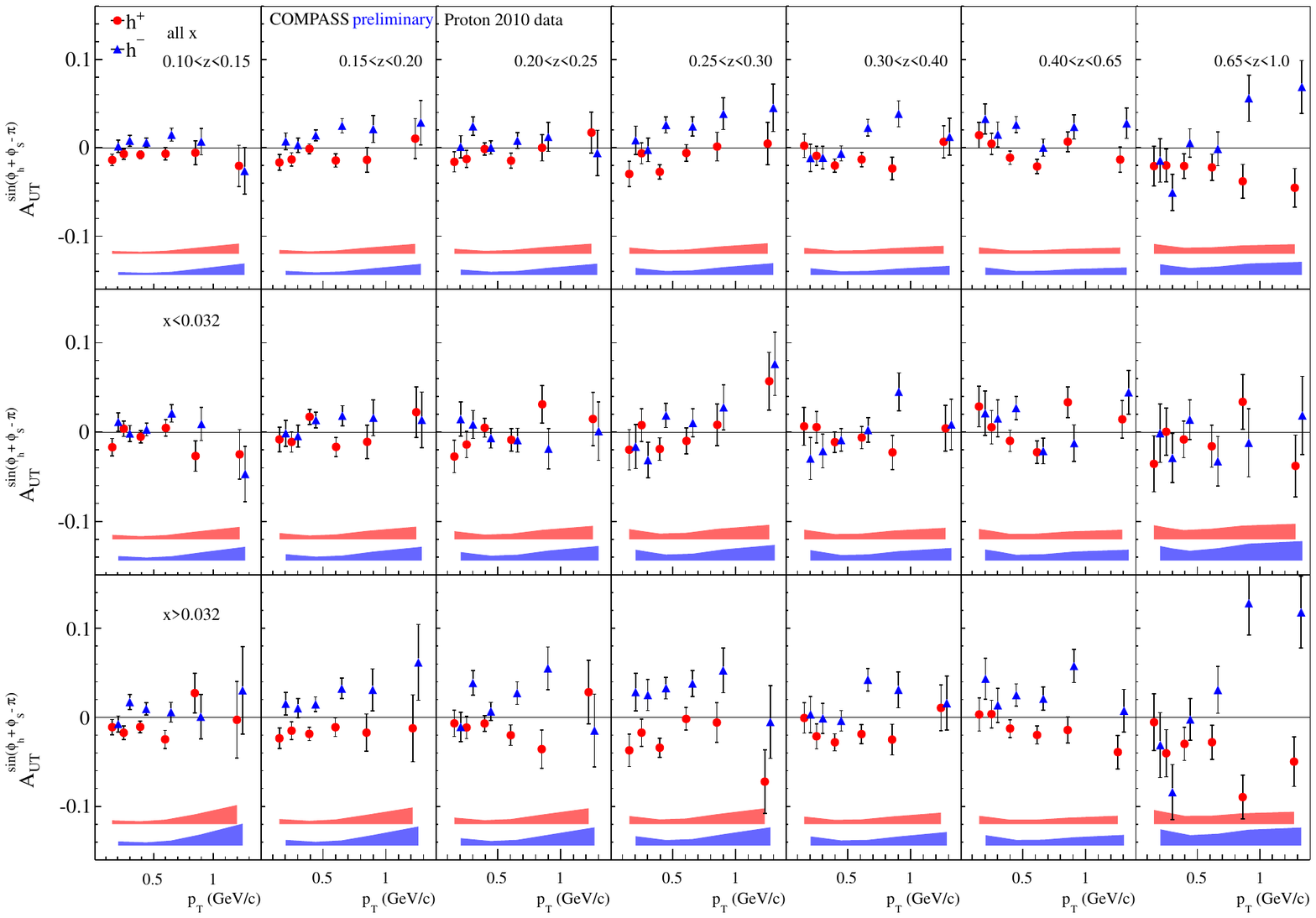}
\vspace*{8pt}
\caption{Collins asymmetry in "3D: $Q^2$-$z$-$x$" (top) and "3D: $x$-$z$-$p_T$" (bottom) configurations. \label{f3}}
\end{figure}

In Fig.~\ref{f3} Collins asymmetry is shown in "3D: $Q^2$-$z$-$x$" (top) and "3D: $x$-$z$-$p_T$" (bottom) grids. Clear "mirrored" behaviour for positive and negative hadron amplitudes is being observed in most of the bins. Amplitudes tend to increase in absolute value with both $z$ and $p_T$.

\begin{figure}[H]
\includegraphics[width=12.8cm]{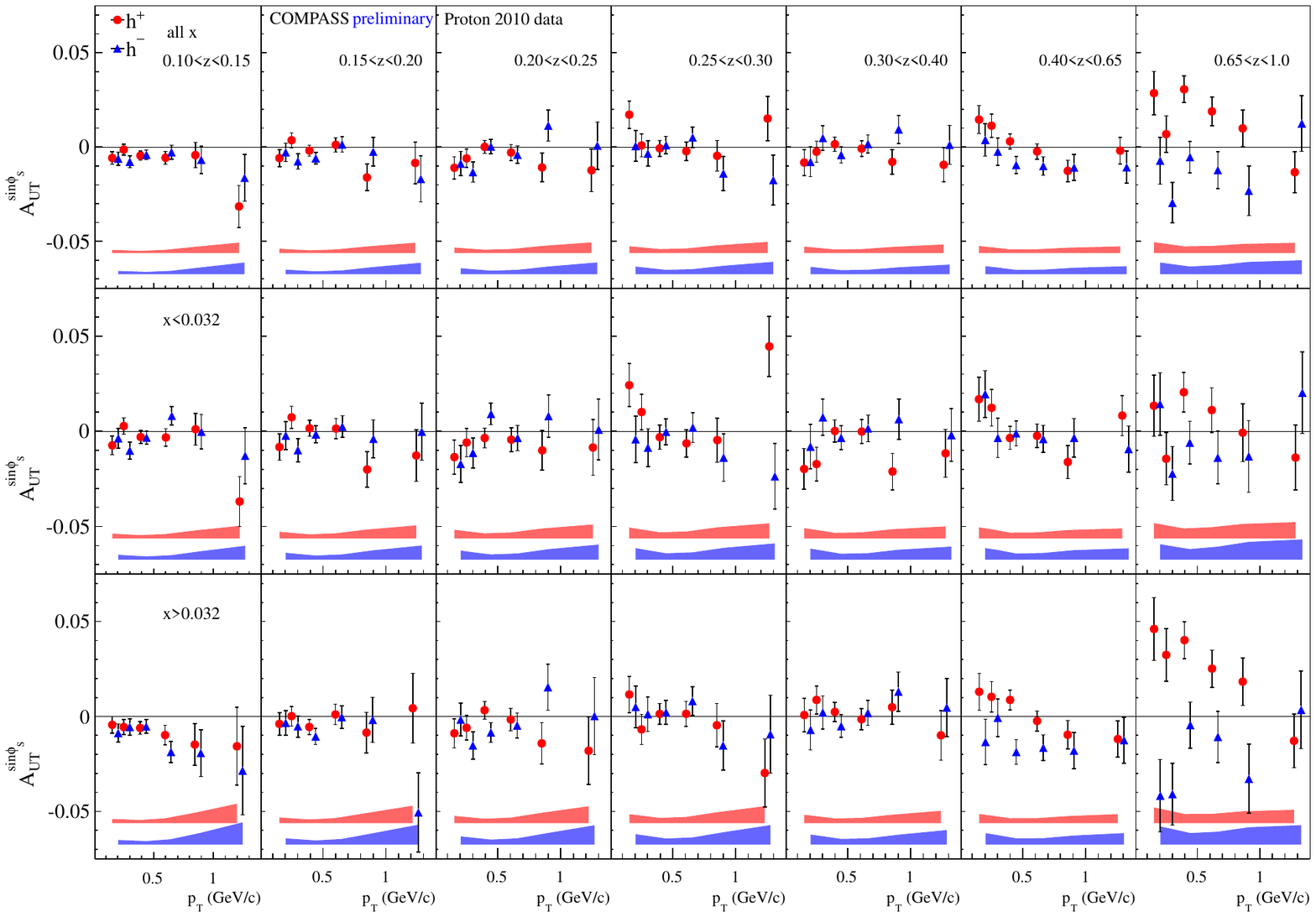}
%\vspace*{8pt}
%\caption{ASiv 3D Q2-z-x. \label{f1}}
%\end{figure}
%\begin{figure}[pb]
%\includegraphics[width=12.8cm]{plots/A_LT_CMPSPr4x4+data_9bin.pdf}
\includegraphics[width=12.8cm]{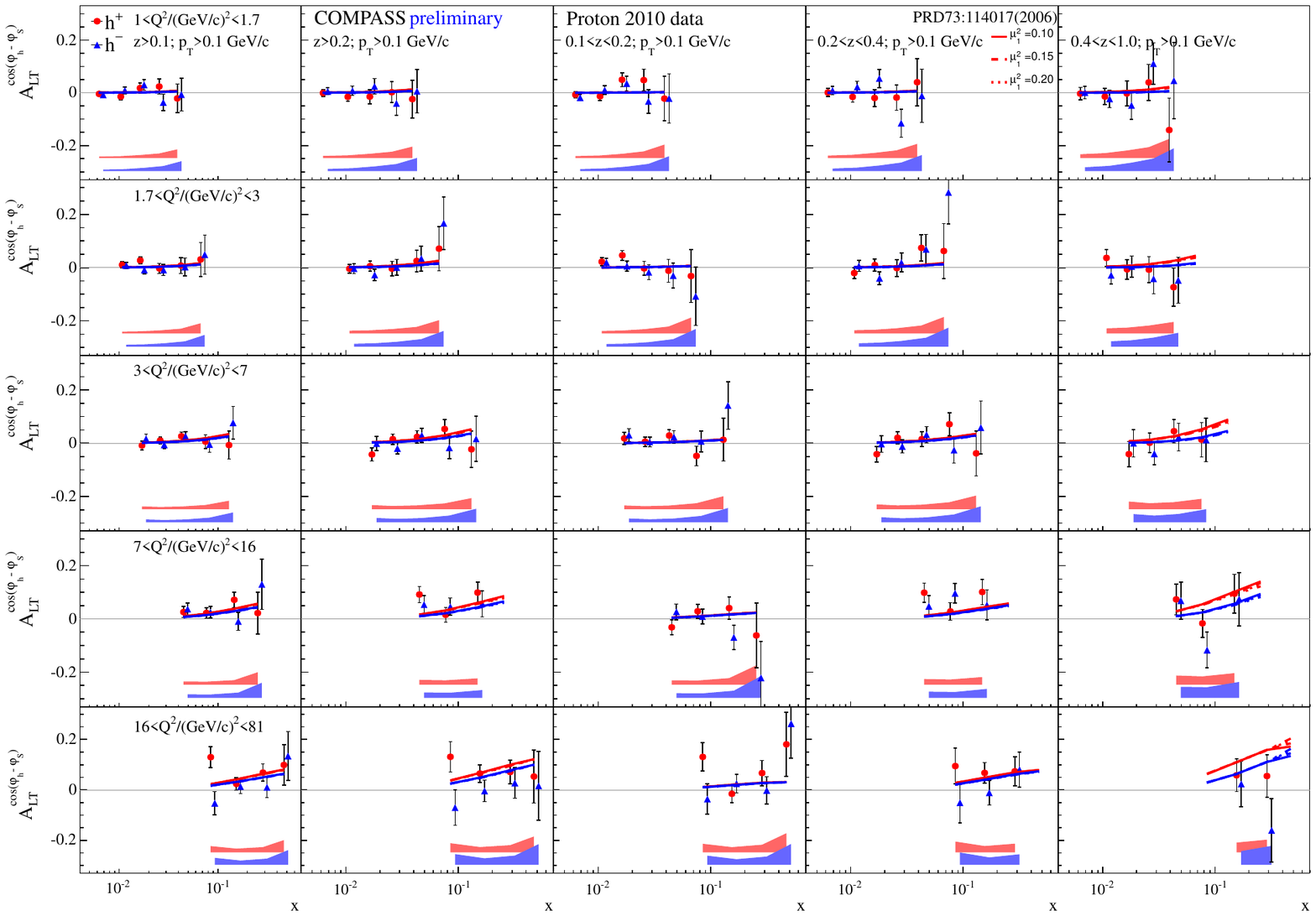}
\caption{Top: $A_{UT}^{\sin (\phi _s )}$ asymmetry in "3D: $x$-$z$-$p_T$". Bottom: $A_{LT}^{\cos (\phiH -\phiS )}$ in "3D: $Q^2$-$z$-$x$" superimposed with theoretical predictions from Ref. 13.
%Ref.~\refcite{Kotzinian:2006dw}
\label{f4}}
\end{figure}

Another SSA which is found to be non-zero at COMPASS is the $A_{UT}^{\sin (\phi _s )}$ term which is presented in Fig.~\ref{f4} (top) in "3D: $x$-$z$-$p_T$" configuration. Here the most
interesting is the large $z$-range were amplitude is measured to be sizable and non zero both for positive and negative hadrons.

The bottom plot in the Fig.~\ref{f4} is dedicated to the $A_{LT}^{\cos (\phiH -\phiS )}$ DSA explored in "3D: $Q^2$-$z$-$x$" grid and superimposed with the theoretical curves from Ref.~\refcite{Kotzinian:2006dw}.
This is the only DSA which appears to be non-zero at COMPASS and the last TSA for which a statistically significant signal has been detected. Remaining four asymmetries are found to be small or compatible with zero within available statistical accuracy and to be in agreement with available predictions \cite{Mao:2014aoa,Mao:2014fma,Lefky:2014eia}.
\section{Conclusions}
The first ever multidimensional extraction of the whole set of target transverse spin dependent azimuthal asymmetries has been done at COMPASS with proton data collected in 2010.
Various multi-differential configurations has been tested exploring $x$:$Q^2$:$z$:$p_T$ phase-space. Particular attention was given to probes of possible $Q^2$-dependence of TSAs, serving a direct input to TMD-evolution related studies.
Several interesting observations have been made studying the results obtained for Sivers, Collins, $A_{LT}^{cos(\phi_h-\phi_S)}$ and $A_{UT}^{sin(\phi_S)}$ asymmetries. Other four asymmetries were found to be compatible with zero within given statistical accuracy.
These results combined with past and future data of other collaborations will give a
unique opportunity to access the whole set of TMD PDFs and test their multi-differential nature and key features.
\\\\
Bakur Parsamyan undertook this work with the support of the ICTP TRIL Programme, Trieste, Italy.
%
%
%References are to be listed in the order cited in the text in Arabic
%numerals.  They should be listed according to the style shown in the
%References. Typeset references in 9 pt roman.
%
%References in the text can be typed in superscripts,
%e.g.: ``$\ldots$ have proven\cite{autbk}\cdash\cite{rvo} that
%this equation $\ldots$'' or after punctuation marks:
%``$\ldots$ in the statement.\cite{rvo}'' This is
%done using LaTeX command: ``$\backslash$cite\{name\}''.
%
%When the reference forms part of the sentence, it should not
%be typed in superscripts, e.g.: ``One can show from
%Ref.~\refcite{autbk} that $\ldots$'', ``See
%Refs.~\refcite{jpap}--\refcite{autbk}, \refcite{rvo}
%and \refcite{pro} for more details.''
%This is done using the LaTeX
%command: ``Ref.$\sim\backslash$refcite\{name\}''.

%\begin{thebibliography}{000} %for 3 digits
%\begin{thebibliography}{00}  %for 2 digits
%

%\clearpage
%

\end{document}